\documentstyle[seceq,epsf,wrapfig,twoside]{ptptex}
\notypesetlogo  %comment in if to eliminate PTPTeX logo
\title{Semi-Leptonic ${\mib B}$ Meson Decays to Excited {\mib D} Mesons\\
in the Covariant Oscillator Quark Model }
\author{%
Masuho {\sc Oda}, Kazunori {\sc Nishimura},$^{*}$\\
Muneyuki {\sc Ishida}$^{**}$ and Shin {\sc Ishida}$^{*}$
}
\inst{%
Faculty of Engineering, Kokushikan University, 
Tokyo 154-8515, Japan\\
$^{*}$Atomic Energy Research Institute, 
College of Science and Technology\\
Nihon University, Tokyo 101-0062, Japan\\ 
$^{**}$Department of Physics, Tokyo Institute of Technology\\
Tokyo 152-8551, Japan
}
\recdate{%
January 31, 2000
}
\abst{%
The spectra and branching ratios of
the weak semi-leptonic $B$ meson decays to the first excited 
$D$ mesons are predicted, taking into account the confined effects of 
quarks using the covariant oscillator quark model (COQM).
In the COQM the same relation between general weak transition form factors
as that in HQET is derived, and the concrete form of the 
Isgur-Wise function
is given with no free parameters. 
Our results are somewhat different from those of other models.
The present experimental data are not sufficient for 
comparison.
}
\begin{document}
\maketitle

\setcounter{tocdepth}{4}

\section{Introduction}

The decay of $B$ mesons has been one of the most important 
topics of high energy physics for many years, and its theoretical
explanation is a matter of great urgency, since now 
a large number of $B$ mesons are produced in $B$-factory experiments.
However, it has been a difficult task to predict 
the spectra and widths of such decays quantitatively, since they are largely 
affected by the confined effects of quarks.
Among many possible channels, an analysis of 
the $B\rightarrow D^*l\bar{\nu}_l$ 
and $B\rightarrow Dl\bar{\nu}_l$ decays is one of the main interests 
in the heavy quark effective theory (HQET).\cite{rf1}
In HQET, all generally independent form factors, 
appearing in the effective meson transition 
current $J_\mu^{B\rightarrow D^*/D}$, 
are represented by one universal form factor (FF) function
(the Isgur-Wise function $\xi (\omega )$), thus leading to 
various FF-relations among them. 
The value of $\xi (\omega )$ at the zero-recoil point
is to be unity $[\xi (1)=1]$ in the heavy quark mass limit
$m_Q\rightarrow\infty$, reflecting
the conserved charge of heavy quark symmetry (HQS).
This function $\xi (\omega )$ describes the confined effects of quarks.
However, HQET and/or HQS  themselves are not able to
predict the concrete form of $\xi (\omega )$, and thus
they cannot describe the FF function and 
the decay spectra in all regions of $q^2$.
For this, it is, in principle, 
necessary to know covariant wave functions (WF)
of mesons for both spin and space-time variables,
and presently we are required to resort 
to some models with a covariant framework.
%The FF function is obtained as an overlapping of initial and 
%final state WF.

We use the covariant oscillator quark model (COQM)\cite{rf2}
in order to estimate the confined effects of quarks.
One of the most important motives for this
model is to describe covariantly the center-of-mass 
motion of hadrons, while preserving the considerable success of
the non-relativistic quark model regarding the static properties of
hadrons. A keystone in COQM for doing this is directly treating 
the squared masses of hadrons, rather than the mass itself, as done 
in conventional approaches. This makes the covariant treatment
simple. The COQM has a long history of development,\cite{rf3} and 
its origin is traced back to the bilocal theory of Yukawa.\cite{rf4}
The COQM has been applied to various problems\cite{rf5} 
with satisfactory results. 
Both of the covariant space-time and spin WF of general meson systems are
determined from the analyses of static problems such as mass spectra\cite{rf51}
and radiative transitions.\cite{rf52}
Recently, Ishida et al.
\cite{rf6,rf7} have studied the weak decays of heavy 
hadrons using this model
and derived the same relations of weak form factors 
for the heavy-to-heavy transition as those derived with HQET.\cite{rf1} 
In addition, our model is also applicable to heavy-to-light
transitions. As a consequence, this model does incorporate
the features of heavy quark symmetry and can be used to
compute the form factors for heavy-to-light transitions as well,
which is beyond the scope of HQET. Actually, in previous
papers we made analyses of the spectra of exclusive 
semi-leptonic\cite{rf7,rf8} decays of $B$-mesons,
of non-leptonic decays of $B$ mesons,\cite{rf9} 
of hadronic weak decays of 
$\Lambda_b$ baryons,\cite{rf10} and of rare radiative decays of 
$B$ mesons\cite{rf11} and $\Lambda_b$ baryons,\cite{rf12} 
leading to encouraging results.
In this paper we extend the application of our model to
weak semi-leptonic decays of $B$ mesons 
to the first excited $D$ mesons, $D^{**}$.
It should be noted that 
the description by HQET of this decay process into excited mesons  
is less reliable than that of decays to ground $D/D^*$ mesons, 
since the approximation in the HQS limit, 
where the momentum of the light degrees of freedom
is ignored in comparison with $m_Q$, is worse in the excited $D$ mesons
than in the ground $D$ mesons.
In contrast with this situation, the description with COQM 
is expected to be reliable, as discussed in 
the third work of Ref. \citen{rf2},
also in this case.

\section{ Model Framework, form factors
and decay width\\
 of ${\mib B}\to ({\mib D}^{*}_2, {\mib D}^{j_q=3/2}_{1},
 {\mib D}^{j_q=1/2}_{1*},
 {\mib D}^{*}_0){\mib l}\bar{\mib \nu}_l$  }

\subsection{Wave functions in COQM}

The general treatment of COQM may be called
the ``boosted $LS$-coupling\\
scheme,"\cite{rf2} 
where the wave functions (WF) are 
tensors in $\tilde U(4) \times O(3,1) $-space and reduce to
those in $SU(2)_{\rm spin} \times O(3)_{\rm orbit} $-space in the
nonrelativistic quark model in the hadron rest frame. The
spinor and space-time portion of the WF
separately satisfy the respective covariant equations,
the Bargmann-Wigner (BW) equation for the former and the
covariant oscillator equation for the latter. The form of
the meson WF has been determined completely
through the analysis of the mass spectra.

In COQM, the meson states are described by bi-local fields
${\Phi_A}^B(x_{1\mu},x_{2\mu}) $, where $x_{1\mu}(x_{2\mu})$ is
the space-time coordinate of the constituent quark (antiquark),
and $A=(a,\alpha)\;(B=(b,\beta))$ describes its flavor and
covariant spinor. Here we write only the positive frequency
part of the relevant ground states:
\begin{equation}
{\Phi_{P,A}}^B(x_{1\mu},x_{2\mu})=e^{iP\cdot X}\; {U(P)_A}^B\;
f_\beta (x_{\mu};P)\;,
\end{equation}
where $U$ and $f_\beta$ are the covariant spinor and internal
space-time WF, respectively, satisfying the
Bargmann-Wigner and oscillator wave equations. The quantity
$x_{\mu} (X_{\mu}) $ is the relative (CM) coordinate,
$x_{\mu}\equiv x_{1\mu}-x_{2\mu}\  
(X_{\mu}\equiv (m_1x_{1\mu}+m_2x_{2\mu})/(m_1+m_2) $, 
where the $m_i$ represent the quark
masses). The function $U$ is given by
\begin{equation}
U(P)=\frac{1}{2 \sqrt 2}\left [(-\gamma_5 P_s(v)+i\gamma_{\mu}
V_{\mu}(v))(1+iv \cdot \gamma)\right ] ,
\end{equation}
where $P_s(V_s)$ represents the pseudoscalar (vector) meson
field, and $v_{\mu} \equiv P_{\mu}/M$ [$P_{\mu}(M)$ is the
four momentum (mass) of the meson]. The function $U$, being
represented by the direct product of quark and antiquark
Dirac spinors with the meson velocity, is reduced to the
non-relativistic Pauli-spin function in the meson rest frame.
The function $f_\beta$ is given\footnote{
In this paper we employ the pure-confining approximation,
neglecting the effect of the one-gluon-exchange potential,
which is expected to be good for the heavy/light-quark meson system,
since this system has comparatively large space extension (see ref. \citen{rf51}).
} by
\begin{equation}
f_\beta (x_{\mu};P)=\frac{\beta}{\pi} \exp\left (-\frac{\beta}{2}
\left (x_{\mu}^2+2(v\cdot x)^2\right ) \right )\ ; \ \ \ 
\beta =\sqrt{\mu K} ,
\end{equation}
where $\mu$ is the reduced mass, and $K$ is the spring constant
of oscillator potential. The quantity $\beta$ represents the inverse ``size''
of WF.

The WF of $D^{**}$ mesons are obtained by operating with  
the creation operator $a_\lambda^\dagger 
= (1/\sqrt{2\beta '})(\beta 'x_\lambda -\partial /\partial x_\lambda )$
on the ground state WF as
\begin{eqnarray}
\Phi_{P'}(x_{1\mu},x_{2\mu})
&=&
\frac{1}{2\sqrt{2}}(\epsilon_{\nu\lambda}^{(h)}(v')i\gamma_\nu
-\epsilon_\lambda (v')\gamma_5)(1+iv'\cdot\gamma )
e^{iP'\cdot X'}\ a_\lambda^\dagger \ 
f_{\beta '} (x_{\mu};P') \nonumber\\
&=&
\frac{1}{2\sqrt{2}}(\epsilon_{\nu\lambda}^{(h)}(v')i\gamma_\nu
-\epsilon_\lambda^{(h)} (v')\gamma_5)(1+iv'\cdot\gamma )\nonumber\\
 & & \times e^{iP'\cdot X'}
\sqrt{2\beta '}(x_\lambda +v'_\lambda v'\cdot x)
f_{\beta '} (x_{\mu};P') ,
\end{eqnarray}
where $\epsilon_{\nu\lambda}^{(h)}(v')\ (\epsilon_\lambda^{(h)}(v'))$
represents the polarization tensor of the final $D^{**}$ meson in 
$^3P_{J=2,1,0}\ (^1P_{J=1})$ states, and $\beta '$ is the inverse size
of the final $D$ meson system.
The Pauli-conjugate of WF is defined by 
$\bar\Phi_{P'}\equiv -\gamma_4\Phi_{P'}^\dagger\gamma_4$.
When the final $D$ mesons are to move in the $z$-direction with
velocity $v'_\mu =(0,0,\omega_3,i\omega )\ 
(\omega_3\equiv\sqrt{\omega^2-1})$, 
the conjugate of the polarization tensors are given by
\begin{eqnarray}
{\rm For}\ ^1P_1,\ \  & \ \ & 
   \tilde\epsilon^{(\pm )}(v')=\frac{\mp 1}{\sqrt{2}}(1,\mp i,0,i0),
           \ \ \
   \tilde\epsilon^{(0)}(v')=(0,0,\omega ,i\omega_3)\nonumber\\
{\rm For}\ ^3P_2,\ \  & \ \ & 
   \tilde\epsilon_{\nu\lambda}^{(\pm 2)}(v')   
            =\tilde\epsilon_{\nu}^{(\pm )}
            \tilde\epsilon_{\lambda}^{(\pm )},  \ \ \ 
   \tilde\epsilon_{\nu\lambda}^{(\pm 1)}(v') =\frac{1}{\sqrt{2}}
           (\tilde\epsilon_{\nu}^{(\pm )}
            \tilde\epsilon_{\lambda}^{(0)}
           +\tilde\epsilon_{\nu}^{(0)}
            \tilde\epsilon_{\lambda}^{(\pm )})  \nonumber\\
 & \ \ & \tilde\epsilon_{\nu\lambda}^{(0)}(v') =\frac{1}{\sqrt{6}}
           (\tilde\epsilon_{\nu}^{(+)}
            \tilde\epsilon_{\lambda}^{(-)}
          +2\tilde\epsilon_{\nu}^{(0)}
            \tilde\epsilon_{\lambda}^{(0)}
           +\tilde\epsilon_{\nu}^{(-)}
            \tilde\epsilon_{\lambda}^{(+)})  \nonumber\\
{\rm For}\ ^3P_1,\ \  & \ \ & 
  \tilde\epsilon_{\nu\lambda}^{(\pm 1)}(v')=\frac{\pm 1}{\sqrt{2}}
           (\tilde\epsilon_{\nu}^{(\pm )}
            \tilde\epsilon_{\lambda}^{(0)}
           -\tilde\epsilon_{\nu}^{(0)}
            \tilde\epsilon_{\lambda}^{(\pm )}) 
  =\frac{1}{\sqrt{2}}\epsilon_{\nu\lambda\alpha\beta}v'_\alpha
   \tilde\epsilon_\beta^{(\pm )} \nonumber\\
 & \ \ & \tilde\epsilon_{\nu\lambda}^{(0)}(v')   =\frac{1}{\sqrt{2}}
           (\tilde\epsilon_{\nu}^{(+)}
            \tilde\epsilon_{\lambda}^{(-)}
           -\tilde\epsilon_{\nu}^{(-)}
            \tilde\epsilon_{\lambda}^{(+)})
  =\frac{1}{\sqrt{2}}\epsilon_{\nu\lambda\alpha\beta}v'_\alpha
   \tilde\epsilon_\beta^{(0)} \nonumber\\
{\rm For}\ ^3P_0,\ \  & \ \  & 
   \tilde\epsilon_{\nu\lambda}(v')   =\frac{1}{\sqrt{3}}
           (\tilde\epsilon_{\nu}^{(+)}
            \tilde\epsilon_{\lambda}^{(-)}
           -\tilde\epsilon_{\nu}^{(0)}
            \tilde\epsilon_{\lambda}^{(0)}
           +\tilde\epsilon_{\nu}^{(-)}
            \tilde\epsilon_{\lambda}^{(+)})
 =-\frac{\delta_{\nu\lambda}+v'_\nu v'_\lambda}{\sqrt{3}}  .
\ \ \ 
\end{eqnarray}

\subsection{Effective weak current}
The effective weak interaction describing the $B\to D^{**}+W$ 
process is given by the covariant overlapping of 
the initial and final meson WF,
\begin{eqnarray}
{\cal L}_W &=& \frac{g}{2\sqrt 2}\ V_{cb}\ \int d^4x_1\int d^4x_2\ 
\langle 
\sqrt{2M'}\bar\Phi_{P'}(X',x)i\gamma_\mu (1+\gamma_5)
\sqrt{2M}\Phi_{P}(X,x)
\rangle W_\mu (x_1),\nonumber
\end{eqnarray}
where we denote the momenta (masses) of the initial $B$
and final $D^{**}$ mesons, respectively, as $P_\mu (M)$
and $P'_\mu (M')$. The momentum of the emitted $W$-boson is denoted as
 $q_\mu$.
The effective weak transition current $J_\mu (=V_\mu +A_\mu )$ 
is defined by the equation  
\begin{eqnarray}
{\cal L}_W &=&  \frac{g}{2\sqrt 2}\ V_{cb}\ \int d^4X\ 
              (1/i)  J_\mu (X)\ W_\mu (X) \nonumber\\
  & \propto & V_{cb}\ (2\pi )^4\delta^{(4)}(P-P'-q)
    (1/i)J_\mu (P',P)\ W_\mu (q),
\end{eqnarray}
where $J_\mu (P',P)\equiv J_\mu (X=0)$.
The overlapping of WF is taken separately in the spinor part and 
the space-time part, and  
the final form of $J_\mu (X)$ is given by  
\begin{eqnarray}
 (1/i) J_\mu (X) &=& \sqrt{2M}\sqrt{2M'}
              O_{\mu\lambda}^{\rm s}\ O_\lambda^{\rm x}\label{eq7} \nonumber\\
 O_{\mu\lambda}^{\rm s} &=& 
   \left\langle
     \frac{-1}{2\sqrt 2}(1+iv'\cdot\gamma )
     (i\gamma_\nu\tilde\epsilon_{\nu\lambda}^{(h)}
      +\gamma_5\tilde\epsilon_{\lambda}^{(h)})i\gamma_\mu (1+\gamma_5)
     \frac{-1}{2\sqrt 2}\gamma_5(1+iv\cdot\gamma )
   \right\rangle  \nonumber \\
  &=& \frac{1}{2}\{
\tilde\epsilon_{\nu\lambda}^{(h)}(   
\varepsilon_{\nu\mu\alpha\beta}v_\alpha v'_\beta 
-(\omega +1)\delta_{\nu\mu}+v'_\nu v_\mu -v'_\mu v_\nu  )
+\tilde\epsilon_{\lambda}^{(h)}( v_\mu +v'_\mu ) \} \nonumber\\
 O_\lambda^{\rm x} &=& 
\int d^4x\ e^{-iP'\cdot X'}f_{\beta '}(x,P')\sqrt{2\beta '}
   (x_\lambda +v'_\lambda (v'\cdot x)) f_{\beta}(x,P)
   \  e^{iP\cdot X}\  e^{-iq\cdot x_1} \nonumber\\
 &=& -i F\times (v_\lambda -\omega v'_\lambda )
     e^{i(P-P'-q)\cdot X}, \\
{\rm where} & &  \nonumber\\
 F  &=& \sqrt{2\beta '}\frac{m_d}{C}
\left\{ -(\beta -\beta ')\frac{M}{M_0}+2\beta\omega\frac{M'}{M'_0}
\right\}\ I_{HO}(\omega ), \nonumber\\ 
C &=& (\beta -\beta ')^2+4\beta\beta '\omega^2\ \ \ \ \  \nonumber\\
  I_{HO}(\omega ) &=& \frac{4\beta\beta '}{\beta +\beta '}
    \frac{1}{\sqrt C}\ {\rm exp}\frac{-m_d^2}{2C}\left[
  (\beta +\beta ')\left\{ \left(\frac{M}{M_0}\right)^2
   +\left(\frac{M'}{M'_0}\right)^2-2\omega\frac{MM'}{M_0M'_0}\right\}
  \right.  \nonumber\\
 & & \ \ \ \ \ \ \ \ \ \ \ \ \ \ \ \ \ \ \left. +2(\omega^2-1)\left\{
\beta\left(\frac{M'}{M'_0}\right)^2
+\beta '\left(\frac{M}{M_0}\right)^2\right\}\right] . \nonumber\\
& & \ \ \ \ \ \ \ \ \ \ \ \ \ \ 
\ \ \ (M_0=m_b+m_d,\ M'_0=m_c+m_d)
\label{eq8}
\end{eqnarray}
The function $F$ here describes the confined effects of 
quarks in the relevant processes.

\subsection{Form factor relation}

The HQET suggests that the angular momentum of 
the light degrees of freedom, ${\mib j}_q$, obtained by the composition
of the orbital angular momentum ${\mib L}$ and the spin of the light quark ${\mib S}_q$ 
as ${\mib j}_q={\mib L}+{\mib S}_q$, is approximately a good quantum 
number, and the states of $D^{**}$ with definite 
$j_q=3/2$ and $1/2$ quantum numbers,
$D^{j_q=3/2}$ and $D^{j_q=1/2}$, are expected to be realized 
in nature. The values of the total $J$ of $j_q=3/2$ states are $J=2,1$ 
and those of 
 $j_q=1/2$ are $J=1,0$. The $J=2,0$ states correspond to the 
$^3P_{2,0}$ states in the boosted $LS$-coupling scheme. 
On the other hand, the $J=1$ states with $j_q=3/2$ and $1/2$  
are represented by the 
superposition of $^3P_1$ and $^1P_1$ states as\cite{rf1}
\begin{eqnarray}
D_1^{j_q=3/2} &=& -\sqrt{\frac{1}{3}}D^{{}^3P_1}
                  -\sqrt{\frac{2}{3}}D^{{}^1P_1},\nonumber \\ 
D_{1*}^{j_q=1/2} &=& \sqrt{\frac{2}{3}}D^{{}^3P_1}
                  -\sqrt{\frac{1}{3}}D^{{}^1P_1}.
\label{eq9}
\end{eqnarray}

By taking this superposition into account, the effective currents
$J_\mu$, Eq.~(\ref{eq7}), for the respective $D^{**}$ mesons 
can be rewritten as 
\begin{eqnarray}
J_\mu^{B\to D_2^*\ \ \ \ \ \ }/\sqrt{MM'} &=& 
  F(\omega ) \tilde\epsilon_{\nu\lambda}^{(h)}v_\lambda
  (\epsilon_{\nu\mu\alpha\beta}v_\alpha v'_\beta 
   -(\omega +1)\delta_{\nu\mu} -v'_\mu v_\nu )  \nonumber\\
J_\mu^{B\to D_1^{j_q=3/2}}/\sqrt{MM'} &=& 
  F(\omega ) \frac{-1}{\sqrt{6}} \left[ \tilde\epsilon\cdot v
  (3v_\mu +(2-\omega )v'_\mu ) \right. \nonumber\\
   & & \left. +(1-\omega^2)\tilde\epsilon_{\mu}
  -(\omega +1)\epsilon_{\mu\lambda\alpha\beta}v_\lambda v'_\alpha 
      \tilde\epsilon_{\beta}  \right]   \nonumber\\
J_\mu^{B\to D_{1^*}^{j_q=1/2}}/\sqrt{MM'} &=& 
  F(\omega ) \frac{1}{\sqrt{3}} \left[ -\tilde\epsilon\cdot v
  (\omega +1 )v'_\mu \right.  \nonumber\\
  & & \left. +(1-\omega^2)\tilde\epsilon_{\mu}
  -(\omega +1)\epsilon_{\mu\lambda\alpha\beta}v_\lambda v'_\alpha 
      \tilde\epsilon_{\beta}  \right]   \nonumber\\
J_\mu^{B\to D_0^*\ \ \ \ \ \ }/\sqrt{MM'} &=& 
  F(\omega ) \frac{1}{\sqrt{3}} (\omega +1)(v_\mu -v'_\mu ) .
\label{eq10}
\end{eqnarray}
The general weak transition form factors are defined by
\begin{eqnarray}
J_\mu^{B\to D_2^*\ \ \ \ \ \ }/\sqrt{MM'} &=& 
  \tilde h \epsilon_{\mu\alpha\beta\gamma}
   \tilde\epsilon_{\alpha\nu} v_\nu (v+v')_\beta (v-v')_\gamma
  -\tilde k  \tilde\epsilon_{\mu\nu} v_\nu \nonumber\\
 & &  + \tilde\epsilon_{\alpha\beta} v_\alpha v_\beta 
    \left[ \tilde b_+ (v+v')_\mu +\tilde b_- (v-v')_\mu 
                                        \right]  \nonumber\\
J_\mu^{B\to D_1^{j_q=3/2}}/\sqrt{MM'} &=& 
   \tilde l_{3/2}  \tilde\epsilon_{\mu}  
  - \tilde\epsilon\cdot v 
    \left[ \tilde c_{3/2+} (v+v')_\mu +\tilde c_{3/2-} 
     (v-v')_\mu\right] \nonumber\\
 & &  -\tilde q_{3/2} \epsilon_{\mu\alpha\beta\gamma}
   \tilde\epsilon_{\alpha} (v+v')_\beta (v-v')_\gamma \nonumber\\
J_\mu^{B\to D_{1^*}^{j_q=1/2}}/\sqrt{MM'} &=& 
   \tilde l_{1/2}  \tilde\epsilon_{\mu}  
  - \tilde\epsilon\cdot v 
    \left[ \tilde c_{1/2+} (v+v')_\mu +\tilde c_{1/2-}
     (v-v')_\mu\right] \nonumber\\
 & &   -\tilde q_{1/2} \epsilon_{\mu\alpha\beta\gamma}
   \tilde\epsilon_{\alpha} (v+v')_\beta (v-v')_\gamma \nonumber\\
J_\mu^{B\to D_0^*\ \ \ \ \ \ }/\sqrt{MM'} &=& 
 \tilde u_{+} (v+v')_\mu +\tilde u_{-} (v-v')_\mu  .
\label{eq11}
\end{eqnarray}
By comparing Eq.~(\ref{eq10}) with Eq.~(\ref{eq11}), we can derive
the following relation between the form factors:
\begin{eqnarray}
\tilde h &=& -\tilde b_+ = \tilde b_- = (1/2) F(\omega ),\qquad 
\tilde k=(\omega +1)F(\omega ). \nonumber\\
\tilde l_{3/2} &=& ((\omega^2-1)/\sqrt{6}) F(\omega ), \nonumber \\ 
  \tilde c_{3/2+} &=& ((5-\omega )/(2\sqrt{6})) F(\omega),\qquad  
      \tilde c_{3/2-}=\tilde q_{3/2}
      = ((1+\omega )/(2\sqrt{6})) F(\omega), \nonumber\\
\tilde l_{1/2} &=& ((1-\omega^2)/\sqrt{3}) F(\omega ), \qquad 
 \tilde c_{1/2+}= -\tilde c_{1/2-} = -\tilde q_{1/2}
 = ((\omega +1)/(2\sqrt{3})) F(\omega )  , \nonumber\\
\tilde u_{+} &=& 0,\qquad  \tilde u_{-} = ((\omega +1)/\sqrt{3}) F(\omega ) .
\label{eq12}
\end{eqnarray}
The above relations are the same as those obtained with 
HQET.\cite{rfiw,rffalk}
 In HQET, the 
$B\rightarrow D^{**}$ current is described by two independent 
Isgur-Wise functions,
$\tau_{3/2}(\omega )$ and $\tau_{1/2}(\omega )$, whose concrete forms
 are not
derivable from HQET.  
In our scheme they are represented  by a single overlapping
$F$ function Eq.~(\ref{eq8}) as 
\begin{eqnarray}
\tau_{3/2}(\omega )=F(\omega ),\qquad 
\tau_{1/2}(\omega )=F(\omega )\ (\omega +1)/\sqrt{3}\ .
\label{eqIW}
\end{eqnarray}
Calculations to derive these equations are given in Appendix A.

\subsection{Helicity amplitude and decay spectra}
Helicity amplitudes $H_h$ are defined by decomposing $J_\mu (X=0)$
into $W$-boson polarization vectors $e_\mu^{(h)}(q)$ as  
\begin{eqnarray}
(1/i) J_\mu (X=0) &=& i(H_+e_\mu^{(+)}(q)+H_-e_\mu^{(-)}(q)+H_0e_\mu^{(0)}(q)
-H_se_\mu^{(s)}(q) ).\ \ \ \ \ \ 
\end{eqnarray}
Thus, $H_h$ is obtained as  
$H_h= \tilde e_\mu^{(h)}(q)\ J_\mu $.
By supposing that the $W$-boson moves in the $-z$ direction with momentum 
$q_\mu =(0,0,-|{\mib q}|, iq_0)=(0,0,-M'\omega_3,i(M-M'\omega ))$ 
in the rest frame of the initial $B$ meson,
the conjugate of the polarization vectors are 
% obtained  (by $\pi$-rotation
%around $y$-axis from normal polarization vector 
%with its momentum in $+z$ direction) as 
given by 
\begin{eqnarray}
 \tilde e_\mu^{(\pm )}(q) &=& 
  \frac{\pm 1}{\sqrt 2}(1,\pm i,0,i0),\ \ \ 
 \tilde e_\mu^{(0)}(q) = 
  \frac{1}{\sqrt{-q^2}}(0,0,-q_0,i|{\mib q}|),\ \ \ 
 \tilde e_\mu^{(s)}(q) =
  \frac{1}{\sqrt{-q^2}}q_\mu .\nonumber
\end{eqnarray}
The actual forms of the helicity amplitudes describing the decay into
the $D^{**}$ mesons in the $^{2S+1}L_J$ state, 
$H_h^{{}^{2S+1}L_J}$, 
are given by  
\begin{eqnarray}
H_\pm^{{}^3P_2} &=& \sqrt{MM'}F\frac{-\omega_3}{\sqrt 2}
(\mp\omega_3+\omega +1),\nonumber\\
       H_0^{{}^3P_2} &=& -\sqrt{\frac{2}{3}} \sqrt{MM'}F   
         (M-M')\frac{\omega_3(\omega +1)}{\sqrt{-q^2}},\nonumber\\
       H_s^{{}^3P_2} &=& -\sqrt{\frac{2}{3}} \sqrt{MM'}F
          (M+M')\frac{\omega_3^2}{\sqrt{-q^2}},\nonumber\\ 
H_\pm^{{}^3P_1} &=& \pm\sqrt{MM'}F\frac{-\omega_3}{\sqrt 2}
(\mp\omega_3+\omega +1),
\ \ \ H_{0,s}^{{}^3P_1} =0\nonumber\\
H_\pm^{{}^1P_1} &=& 0,\qquad 
H_0^{{}^1P_1} = \sqrt{MM'}F
(M+M')\frac{\omega_3^2}{\sqrt{-q^2}},\nonumber \\ 
  H_s^{{}^1P_1} &=& \sqrt{MM'}F
(M-M')\frac{\omega_3(\omega +1)}{\sqrt{-q^2}},\nonumber\\ 
H_\pm^{{}^3P_0} &=& 0,
\qquad  H_0^{{}^3P_0} = \frac{1}{\sqrt 3} \sqrt{MM'}F
(M-M')\frac{\omega_3(\omega +1)}{\sqrt{-q^2}},\nonumber \\ 
  H_s^{{}^3P_0} &=& \frac{1}{\sqrt 3} \sqrt{MM'}F
(M+M')\frac{\omega_3^2}{\sqrt{-q^2}}.
%\label{eq1}
\end{eqnarray}

By taking into account the superposition of Eq.~(\ref{eq9}), 
the helicity amplitudes describing the decays 
into the states with definite $j_q$ quantum numbers are given by 
\begin{eqnarray}
H_\pm^{D_1^{j_q=3/2}} &=& -\sqrt{\frac{1}{3}}H_\pm^{{}^3P_1},\qquad  
H_{0,s}^{D_1^{j_q=3/2}} = -\sqrt{\frac{2}{3}}H_{0,s}^{{}^1P_1},\nonumber\\
H_\pm^{D_{1*}^{j_q=1/2}} &=& \sqrt{\frac{2}{3}}H_\pm^{{}^3P_1},\qquad  
H_{0,s}^{D_{1*}^{j_q=1/2}} = -\sqrt{\frac{1}{3}}H_{0,s}^{{}^1P_1}.
\end{eqnarray}

Decay spectra are obtained by using these $H_h$ through the usual 
procedure:
\begin{eqnarray}
\frac{d\Gamma}{dq^2} &=& \frac{|V_{cb}|^2G_F^2}{192\pi^3M^2}|{\mib p}'|
\left( 1+\frac{m_l^2}{q^2}\right)^2(-q^2)\nonumber\\
& & \times \left[ \left( 2-\frac{m_l^2}{q^2} \right)
\left( |H_+|^2+|H_-|^2+|H_0|^2 \right)-3\frac{m_l^2}{q^2}|H_s|^2
\right] ,
\end{eqnarray}
where $m_l$ is the mass of the lepton and  
$G_F\ (=1.166392\times 10^{-5}$ GeV$^{-2})$ is the Fermi coupling constant. 

The results for the respective $D^{**}$ mesons are given by
\begin{eqnarray} 
\frac{d\Gamma^{D_2^*} }{dq^2}
   &=& \frac{|V_{cb}|^2G_F^2MM'^2}{96\pi^3}\omega_3
\left( 1+\frac{m_l^2}{q^2}\right)^2 F(\omega )^2 
 \  \left[ \left( 1-\frac{m_l^2}{2q^2} \right)
 \omega_3^2 (\omega+1)^2  \right. \nonumber\\
&\times&  \left. \left\{  2\omega \frac{1+r^2-2r\omega }{\omega +1}
  +\frac{2}{3}(1-r)^2  \right\}
  +m_l^2 \frac{\omega_3^4(1+r)^2}{-q^2}  \right] .\  \\
\frac{d\Gamma^{D_1^{j_q=3/2}} }{dq^2}
   &=& \frac{|V_{cb}|^2G_F^2MM'^2}{96\pi^3}\omega_3
\left( 1+\frac{m_l^2}{q^2}\right)^2 F(\omega )^2 (\omega+1)^2 
  \left[ \left( 1-\frac{m_l^2}{2q^2} \right)
 \frac{2}{3} (\omega-1)^2  \right. \nonumber\\
&\times& \left. \left\{  \omega \frac{1+r^2-2r\omega }{\omega -1}
  +(1+r)^2  \right\}
+m_l^2 \frac{\omega_3^2 (1-r)^2}{-q^2}\right] . \\
\frac{d\Gamma^{D_{1^*}^{j_q=1/2}} }{dq^2}
   &=& \frac{|V_{cb}|^2G_F^2MM'^2}{96\pi^3}\omega_3
\left( 1+\frac{m_l^2}{q^2}\right)^2 F(\omega )^2 \frac{(\omega +1)^2}{3}
 \left[ \left( 1-\frac{m_l^2}{2q^2} \right)
 (\omega-1)^2  \right.\nonumber\\
& \times & \left. \left\{  4\omega \frac{1+r^2-2r\omega }{\omega -1}
  +(1+r)^2  \right\}
+m_l^2 \frac{3\omega_3^2 (1-r)^2}{-2q^2}\right] . \\
\frac{d\Gamma^{D_{0}^*} }{dq^2}
   &=& \frac{|V_{cb}|^2G_F^2MM'^2}{96\pi^3}\omega_3
\left( 1+\frac{m_l^2}{q^2}\right)^2 F(\omega )^2 \frac{(\omega +1)^2}{3}
\nonumber\\
 & \times &  \left[ \left( 1-\frac{m_l^2}{2q^2} \right)
\left\{  \omega_3^2 (1-r)^2  \right\}
+m_l^2 \frac{3(\omega -1)^2 (1+r)^2}{-2q^2}\right] , 
\end{eqnarray}
where $\omega_3\equiv\sqrt{\omega^2-1}$ and $r\equiv M'/M$
($M'$ being the mass of the final $D^{**}$ meson).

By integrating the decay spectra over $q^2$, we obtain the decay width.
The physical region of $-q^2$ is $m_l^2\leq -q^2\leq (M-M')^2$.
Large (Small) values of $-q^2$ correspond to the non-relativistic (relativistic)
region, and 
its maximum (minimum) value corresponds to zero-recoil $\omega =1$
(maximum-recoil
$\omega =\omega_{\rm max}=(M^2+M'^2-m_l^2)/(2MM'))$.

\subsection{Values of the parameters}
Our scheme contains only two  parameters,
$m_q$ and $K$, related to the quark structure of mesons.
We examine our problem for the following two 
choices of the numerical values:

Case A:\ \ The $m_q$ are simply determined by
$M_V=m_q+m_{\bar{q}}$, $M_V$ being the masses of 
relevant vector mesons, $B^*$ and $D^*$, in the ground $S$-wave states.
The value of $K$ is assumed to be universal,\cite{rf13} that is,  
independent of flavor-contents
of mesons, and determined from 
the Regge slope of the $\rho$ meson trajectory,
$\Omega^{\rm exp}=1.14$ GeV, by the relation
 $\Omega =\sqrt{32m_nK}$.

Case B:\ \ They are determined from recent analyses 
of mass spectra, including effects due to the 
color Coulomb force.\cite{rf14}
In this case, the mixing of the ground 1S-states with excited 2S-states
is shown to be a few percent
for $D$ or $D^*$ and $B$ in the amplitudes,
and thus its effects seem to be negligible.

The actual values of the $m_q$ and $\beta$ for the respective systems 
are listed in Table I. 
\begin{table}[t]
\caption{The adopted values of quark masses  $m_q$ and 
of the inverse size $\beta$.  
Case A:  
$K=0.106$ GeV${}^3$ is taken to be universal.  
Case B: $K$ is determined from the mass
spectra as $K=$
0.0679 and 0.0619 GeV${}^3$, respectively,
for $D^{**}$ and $B$ mesons.
%In Case B the size of mesons becomes almost equal.
}
\begin{center}
\begin{tabular}{|l|c|c|c||c|c|}
\hline
  & $m_n$ & $m_c$ & $m_b$  & $\beta_{D/D^*}$ & $\beta_B$ \\
\hline
Case A & 0.384 GeV & 1.62 GeV & 4.94 GeV  
                                &  0.181 GeV${}^2$ & 0.194 GeV${}^2$ \\
\hline
Case B & 0.400 GeV & 1.70 GeV & 5.00 GeV  
                                &  0.148 GeV${}^2$ & 0.151 GeV${}^2$ \\
\hline
\end{tabular}
\end{center}
\end{table}

The values of $|V_{cb}|$ and the masses and 
the lifetimes of $\bar B^0$ and $B^\pm$ mesons 
are taken from ref. \citen{rf15}. The masses of $D^{**}$ are from
ref. \citen{rf16}:
\begin{eqnarray}  
|V_{cb}| &=& 0.0395;\qquad  M_B = 5.279\ {\rm GeV},\qquad \
\tau_{B}= 1.6\ {\rm ps},\qquad 
M'_{D_2^*} = 2.459\ {\rm GeV},\nonumber \\ 
& & M'_{D_1^{(j_q=3/2)}} = M'_{D_{1*}^{(j_q=1/2)}} = 2.422\ {\rm GeV},\qquad 
M'_{D_0^*} = 2.36\ {\rm GeV}.
\end{eqnarray}

\section{Results and conclusion}

\begin{wrapfigure}{c}{14cm}
  \epsfysize=6.cm
  \centerline{\epsffile{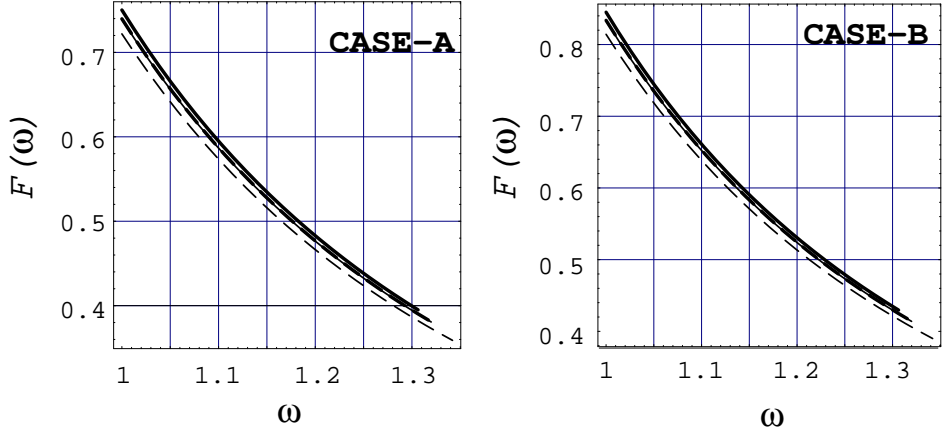}}
%  \figurebox{15.cm}{12cm}
  \caption{$F$ functions of  $B\rightarrow D_2^*$ (thick solid line), 
          $B\rightarrow D_{1}^{j_q=3/2}$ (thick dashed line),
 $B\rightarrow D_{1*}^{j_q=1/2}$ (thin solid line)
           and $B\rightarrow D_0^*$ (thin dashed line) transitions:
           The left figure is for case A, and right figure is for case B. 
           The four lines almost coincide and are hard to be 
           discriminated from each other
                   in the figure.
           The physical regions of $-q^2$, $(M-M')^2\geq -q^2 \geq m_l^2$, 
           correspond to those of $\omega$,
           $1\leq \omega \leq 1.306,1.319,1.319,1.342$, 
           respectively, where we use $m_l$ as the muon mass.
 }
  \label{fig1}
\end{wrapfigure}
First, we display the $-q^2$ dependence of the form factor function $F$
for the $B\to (D_2^*,D_{1}^{j_q=3/2},D_{1*}^{j_q=1/2},D_0^*)$ 
transitions in Fig. 1.
The mass of $D_{1}^{j_q=3/2}$ is taken to be equal to that of
$D_{1*}^{j_q=1/2}$, and thus the corresponding $F$ functions become common.
These functions are almost identical for all these transitions.

\begin{wrapfigure}{c}{14cm}
  \epsfysize=6.cm
 \centerline{\epsffile{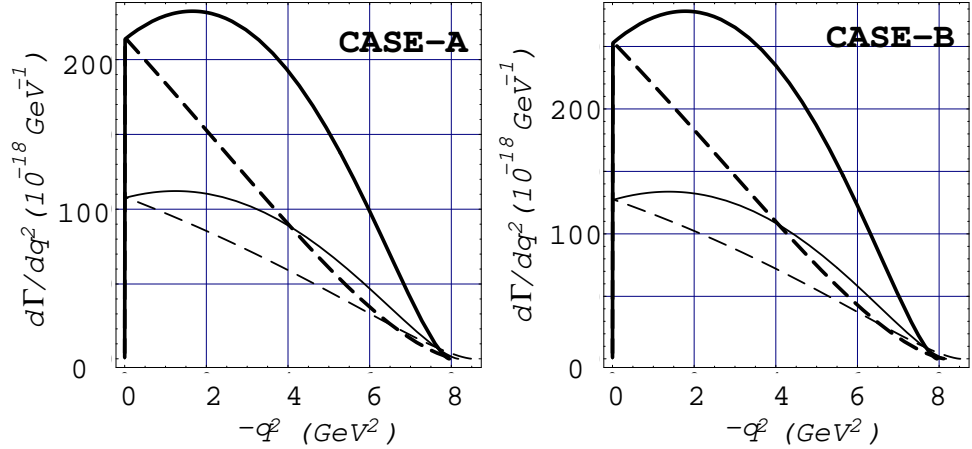}}
%  \figurebox{15.cm}{12cm}
  \caption{Predicted spectra $d\Gamma /dq^2\ (10^{-18}$GeV$^{-1})$
of  $B\rightarrow D_2^*$ (thick solid line), 
          $B\rightarrow D_{1}^{j_q=3/2}$ (thick dashed line),
 $B\rightarrow D_{1*}^{j_q=1/2}$ (thin solid line)
           and $B\rightarrow D_0^*$ (thin dashed line) transitions:
           The left figure is for case A, and the right figure is for case B. 
 }
  \label{fig2}
\end{wrapfigure}
By using these $F$, we can predict the decay spectra,
given in Fig. 2.  
As can be seen in this figure, 
the differential widths in relativistic region
 are much larger than those in non-relativistic region,
since the decay amplitude vanishes at the zero-recoil point due to the on-shell condition
of the final $D^{**}$ mesons,\footnote{
The $P$-wave WF of $D^{**}$ includes the factor $x_\mu$, and thus 
the overlapping
of the initial and final space-time WF becomes proportional 
to $v_\mu$ or $v'_\mu$.
At zero-recoil point $v_\mu$ becomes equal to $v'_\mu$, 
and their contractions to 
the polarization vectors of the $D^{**}$ mesons  vanish 
because of Lorentz condition.
}
and the widths in the non-relativistic region near zero-recoil point
are suppressed comparatively to those in relativistic region.
As discussed in the appendix, in HQET, effectively the same spinor WF as
those in COQM 
are used for the relevant
processes with no clear reason, and accordingly the models basing on HQET also predicts
the comparatively larger differential widths in relativistic region.  
This suggests that 
the quantitative estimation of the quark confined effects 
is crucially important to estimate  the decay widths. 

By integrating out this spectra, 
the theoretical branching ratios $Br_{\rm th}$
are obtained and given in Table II. There they  
are compared with the experimental $Br_{\rm exp}$, 
together with the predictions of
the other theoretical 
models (ISGW model,\cite{rfisgw} CNP model,\cite{rfcnp}
and SISM model\cite{rfsism}). 
Experimental values of Br., $Br_{\rm exp}$,
 for  $\bar B\to D_{1*}^{j_q=1/2}l\bar\nu_l$
and $\bar B\to D_0^*l\bar\nu_l$ have not yet been obtained.
The present  $Br_{\rm exp}$ have large errors
 and seem to be mutually inconsistent.
Our predicted values of $Br_{\rm th}$ are somewhat
different from the other models.
Future experiments will select the correct model.

\begin{table}[t]
\caption{Predicted branching ratios $Br_{\rm th}$  compared with 
experimental values. $Br_{\rm th}$ 
without (with) brackets corresponds to case A (B). 
The lifetime of the $B$ meson is taken as 
$\tau_{B}=1.6$ ps.
The experimental values of Br., $Br_{\rm exp}$, have not yet been obtained 
for the decays 
 $\bar B\to D_{1*}^{j_q=1/2}l\bar\nu_l$ and 
 $\bar B\to D_{0}^*l\bar\nu_l$.
The ISGW model uses the $LS$ coupling scheme,
and the values given in the columns for
 $\bar B\to D_1^{j_q=3/2}l\bar\nu_l$ and
 $\bar B\to D_{1*}^{j_q=1/2}l\bar\nu_l$
are those for $^1P_1$ and $^3P_1$ states, respectively.
}
\begin{center}
\begin{tabular}{|l|c|c|c|c|}
\hline
Br. & $\bar B\to D_2^*l\bar\nu_l$ & $\bar B\to D_1^{j_q=3/2}l\bar\nu_l$ 
  & $\bar B\to D_{1*}^{j_q=1/2}l\bar\nu_l$ & $\bar B\to D_0^*l\bar\nu_l$ \\
\hline
Exp.\ \ \  &  &  &  &  \\
\ \ OPAL & 2.3$\pm$0.9\% ($\bar B^0$)
        & 2.1$\pm$0.9\% ($\bar B^0$)  &  & \\
 & 0.9$\pm$0.4\% ($B^-$)  &  2.1$\pm$0.7\% ($B^-$) &  &  \\
\ \ CLEO & $<$ 1.0\% ($B^-$)
             &  0.49$\pm$0.14\% ($B^-$) &  &  \\
\ \ ALEPH & $<$ 0.15$\sim$0.20\% 
                  & 0.74$\pm$0.16\%  &  &  \\
\hline
Theor.\ \ \  &  &  &  &  \\
\ \ COQM & 0.31(0.38)\%  & 0.19(0.22)\%  & 0.15(0.18)\% 
 & 0.11(0.14)\%  \\
\ \ ISGW & 0.13\% & (0.57\%  for $^1P_1$) 
 & (0.30\%  for $^3P_1$) & 0.14\%  \\
\ \ CNP & 0.24\%  & 0.12\%  & 0.09\%  & 0.07\%  \\
\ \ SISM &0.12\%  & 0.087\%  & 0.036\%  & 0.027\%  \\
\hline
\end{tabular}
\end{center}
\end{table}

Finally, we would like to remark on 
the treatment of $B$ decay with HQET in comparison to COQM.
For describing the decay to the ground and the first excited $D$ mesons,
the three independent Isgur-Wise functions, $\xi (\omega )$, 
$\tau_{3/2}(\omega )$ and $\tau_{1/2}(\omega )$, which are not
derivable only in the general framework of HQET, are necessary.
In COQM, these Isgur-Wise functions are predicted with no free parameters.
Heavy-to-light transitions, 
such as $B\rightarrow\rho$, also cannot be  
described with HQET, while in  
the case of COQM
they can be treated on the same footing as heavy-to-heavy
transitions. 
The treatment using HQET is considered to become less reliable,
in principle, for more excited states,
since the approximation of the HQS limit,  $m_{lq}\ll m_Q$ ($m_{lq}$ is
the mass of the light degrees of freedom),
becomes worse for increasingly excited states.
In COQM the excited states can be treated on the same footing as for 
the ground states.

In this paper we have analyzed the semi-leptonic $B$ meson decays to
the first excited $D$ mesons in the framework of COQM.
All the parameters in COQM have been determined by the analyses 
of mass spectra, and thus the results in this paper are pure predictions,
with no free parameters. 
The framework of COQM has been found to be very effective
in describing general weak decays of ground-to-ground transitions. 
Whether it
is also effective for describing the decay of ground-to-excited 
transitions is a very interesting question. 
However, the present experimental data cannot answer this question.

\acknowledgements

The authors are grateful to Dr. K. Yamada  for informing us of 
his results on the mass spectra. One of the authors, 
M. I, would like to express his sincere gratitude to professor
M. Oka for his hearty encouragement
and also to the members of the nuclear physics  groups 
at the Tokyo Institute of Technology 
for financial support. The authors thank Doctor T. Ishida for his useful suggestions.

\appendix

\section{Derivation of the Form Factor Relation and\\ 
Comparison of Our Spinor Overlapping Calculation with HQET}

Here we compare our overlapping calculation of BW spinors with that
using HQET by Falk et al.\cite{rffalk}
In Ref. \citen{rffalk}, the WF for the final
 mesons with $j_q=1/2$ are taken as  
\begin{eqnarray}
(1-iv'\cdot\gamma )\ \ \ \ \ \ \ \ \ \ \ \ &\ \  &{\rm for}\ \ \ \ \ D_0^* \nonumber\\
\gamma_5(i\tilde\epsilon\cdot\gamma )
(1-iv'\cdot\gamma )\ \ \ \ \ \  &\ \  &{\rm for}\ \ \ \ \ D_{1^*}^{j_q=1/2} ,
\label{eq1/2}
\end{eqnarray}
where we omit the overall normalization factor $1/2\sqrt 2$.
The WF of the final mesons with $j_q=3/2$ are taken as 
\begin{eqnarray}
\sqrt{\frac{3}{2}}\gamma_5\left[ \tilde\epsilon_\lambda 
+\frac{1}{3}(i\gamma_\lambda +v'_\lambda )i\tilde\epsilon\cdot\gamma
    \right](1-iv'\cdot\gamma )
\ \ \ \ \ &\ \ & {\rm for}\ \ \ \ \  D_{1}^{j_q=3/2}  \nonumber\\
-i\gamma_\nu  \tilde\epsilon_{\lambda\nu} (1-iv'\cdot\gamma )\ \ \ 
\ \ \ \ \ \ \ \ \ \ \ \ \ \ \ \ \ \ \ \ \ \ \ \ \ \ \ &\ \ &
{\rm for}\ \ \ \ \  D_2^* .
\label{eq3/2}
\end{eqnarray}
In the overlapping calculation of the initial $B$ and final $D^{**}$ mesons,
the suffix $\lambda$ in Eq.~(\ref{eq3/2}) is contracted with the velocity
of the initial $B$ meson, $v_\lambda$, for no clear reason,
and Eq.~(\ref{eq3/2}) is rewritten as
\begin{eqnarray}
\sqrt{\frac{3}{2}}\gamma_5\left[ \tilde\epsilon\cdot v 
+\frac{1}{3}(iv\cdot\gamma -\omega )i\tilde\epsilon\cdot\gamma
    \right](1-iv'\cdot\gamma )  \ \ \ \  
 &\ \ & {\rm for}\ \ \ \ \   D_{1}^{j_q=3/2}   \nonumber\\
-i\gamma_\nu  \tilde\epsilon_{\lambda\nu} (1-iv'\cdot\gamma )v_\lambda 
\ \ \ \ \ \ \ \ \ \ \ \ \ \ 
&\ \ & {\rm for}\ \ \ \ \ D_2^* .
\label{eq3/2v}
\end{eqnarray}

In COQM the overlapping of space-time WF,  $O_\lambda^x$,
 is proportional to
$(v_\lambda -\omega v'_\lambda )$, as shown in Eq.~(\ref{eq8}).
Contraction of the second term $\omega v'_\lambda$ 
with the spinor overlapping  $O_{\mu\lambda}^s$ vanishes, due to
the on-shell condition of the final $D^{**}$ mesons, and the contraction
of the $v_\lambda$ term only contributes to the effective current $J_\mu$.
Thus, the above contraction of $v_\lambda$ in HQET 
is naturally explained in our scheme.

By including this contraction with $v_\lambda$, 
in the overlapping calculation Eq.~(\ref{eq8})
of COQM, the WF for the final
 mesons with $j_q=1/2$ are taken as
\begin{eqnarray}
(1 &+& iv'\cdot\gamma )(-i\gamma_\nu )
\frac{\delta_{\nu\lambda}+v'_\nu v'_\lambda}{-\sqrt 3}
(v_\lambda -\omega v'_\lambda )  
 = 
\frac{iv\cdot\gamma +\omega }{\sqrt 3}(1-iv'\cdot\gamma )\nonumber\\
 &=&  \frac{1 +\omega }{\sqrt 3}(1-iv'\cdot\gamma )
 \ \ \ \ {\rm for}\ \ \  D_0^*   \ ,
\label{eqcoqm0}
\end{eqnarray}
where in the last equality the term $iv\cdot\gamma$ in the left hand side 
is replaced by 1, because of the on-shell factor  $(1+iv\cdot\gamma )$
of the initial $B$ meson spinor WF.
Equation (\ref{eqcoqm0}) coincides with the first equation of (\ref{eq1/2})
except for the factor $(1+\omega )/\sqrt{3}$.

Similarly, we have 
\begin{eqnarray}
(1 &+& iv'\cdot\gamma )\left[ (-i\gamma_\nu\frac{1}{\sqrt 2}
\epsilon_{\nu\lambda\alpha\beta}v'_\alpha\tilde\epsilon_\beta )
\sqrt{\frac{2}{3}}+(-\gamma_5\tilde\epsilon_\lambda )
(-\sqrt{\frac{1}{3}})   \right](v_\lambda -\omega v'_\lambda )\nonumber\\
&=& -\frac{1}{\sqrt{3}}[\epsilon_{\nu\lambda\alpha\beta}
i\gamma_\nu v_\lambda v'_\alpha \tilde\epsilon_\beta 
-\gamma_5\tilde\epsilon\cdot v](1-iv'\cdot\gamma ) \nonumber\\
 & = & 
\frac{1+\omega }{\sqrt{3}} \gamma_5 i\tilde\epsilon\cdot\gamma 
(1-iv'\cdot\gamma ) \ \ \ \ \ \ {\rm for}\ \ \  D_{1^*}^{j_q=1/2} 
\label{eqcoqm1*}
\end{eqnarray}
where in the last equality we have used the formula 
$\epsilon_{\nu\lambda\alpha\beta}i\gamma_\nu 
=-i\gamma_5\gamma_\lambda\gamma_\alpha\gamma_\beta 
+i\gamma_5\gamma_\lambda\delta_{\alpha\beta}
+i\gamma_5\gamma_\beta\delta_{\lambda\alpha}
-i\gamma_5\gamma_\alpha\delta_{\lambda\beta}$.
Equation (\ref{eqcoqm1*})
coincides with the second equation of (\ref{eq1/2}),
except for the factor $(1+\omega )/\sqrt{3}$. 
The conventional Isgur-Wise function $\tau_{1/2}$ is defined in terms of
WF Eq.~(\ref{eq1/2}).
Thus, $\tau_{1/2}$
is represented by our $F$ function as 
\begin{eqnarray}
\tau_{1/2} &=& F(\omega )(\omega +1)/\sqrt{3}.
\label{eqtau12}
\end{eqnarray}

Similarly, by including the contraction with $v_\lambda$
mentioned above, the WF for the final mesons with $j_q=3/2$ are
given by
\begin{eqnarray}
(1&+&iv'\cdot\gamma )\left[ (-i\gamma_\nu\frac{1}{\sqrt 2}
\epsilon_{\nu\lambda\alpha\beta}v'_\alpha\tilde\epsilon_\beta )
(-\sqrt{\frac{1}{3}})+(-\gamma_5\tilde\epsilon_\lambda )
(-\sqrt{\frac{2}{3}})   \right](v_\lambda -\omega v'_\lambda )\nonumber\\
 &=&\sqrt{\frac{3}{2}}\gamma_5\left[
\tilde\epsilon\cdot v +\frac{iv\cdot\gamma -\omega}{3}
i\tilde\epsilon\cdot\gamma \right]  (1-iv'\cdot\gamma ) 
\ \ \ \ {\rm for}\ \ \  D_{1}^{j_q=3/2} ,  
\label{eqcoqm1}   \\
(1 &+& iv'\cdot\gamma )(-i\gamma_\nu )
\tilde\epsilon_{\nu\lambda}(v_\lambda -\omega v'_\lambda )
 =  -i\gamma_\nu \tilde\epsilon_{\nu\lambda}(1-iv'\cdot\gamma 
)v_\lambda \ \ \ \ {\rm for}\ \ \  D_{2}^{*} \ \ .
\label{eqcoqm2}
\end{eqnarray}
These are exactly the same as Eq.~(\ref{eq3/2v}) in HQET,
and the Isgur-Wise function $\tau_{3/2}$ is given as 
\begin{eqnarray}
\tau_{3/2} &=& F(\omega ).
\label{eqtau32}
\end{eqnarray}
Equations (\ref{eqtau12}) and (\ref{eqtau32}) are Eq.~(\ref{eqIW})
in the text.

The essential point for obtaining the same form factor relation
in COQM as in HQET is use of the BW spinor as spin WF,
which is equivalent to direct product of the ``free'' Dirac
spinors of constituent quarks and antiquarks. On the other hand,
in HQET only the heavy quark is argued to be on shell,
while the momentum of the light quark (or light degrees of freedom)
is not defined clearly. 
However, as discussed previously,\cite{rf2}
fixing the velocity of the heavy quark, $v_{Q\mu}$, 
to be equal to the meson velocity,
 $v_\mu$, necessarily leads to 
 the velocity of the light quark, $v_{q\mu}$, 
also being equal to $v_\mu$, 
which implies that 
the light quark also is on shell.
Accordingly, the BW spinor WF is also implicitly used in HQET.

Finally, we comment on the SISM model.\cite{rfsism}
In this model,
to obtain the same form factor relation, the spinor WF 
given by Falk et al. is applied. Their 
 Isgur-Wise functions $\tau_{1/2}$ and  $\tau_{3/2}$,
calculated using the overlapping of the initial and final WF in NRQM,  
are argued to vanish at the zero-recoil point, 
$\omega =1$, since the initial and final WF are orthogonal in the HQS limit
in this case. However, this argument is not correct.
In COQM, the overlapping of space-time WF, $O_\lambda^x$,
is given by  $O_\lambda^x \propto F(\omega )
(v_\lambda -\omega v'_\lambda)$. At the zero-recoil point, where
$v_\lambda =v'_\lambda$ and $\omega =1$, $O_\lambda^x$ vanishes
due to the factor $(v_\lambda -\omega v'_\lambda)$.
However, the ``Isgur-Wise function'' $F(\omega )$ (or  
$F(\omega )(1+\omega )/\sqrt{3}$) does not vanish.
The factor $(v_\lambda -\omega v'_\lambda)$ already appeared for
the spinor overlapping calculation in HQET, as was discused above,
and this seems to imply in the calculation of the SISM model that the factor 
$(v_\lambda -\omega v'_\lambda)$ is used twice. 
This may be the reason why their branching ratios are 
much smaller than those obtained with the ISGW model,
although both models are based on NRQM.

\end{document}